\documentclass[manuscript]{aastex}
\usepackage{natbib}
\usepackage{hyperref}
\usepackage{graphicx}
\usepackage{bm}
\usepackage{amsmath}
\usepackage{amsfonts}
\usepackage{threeparttable}
\usepackage{float}
\usepackage{morefloats}

\begin{document}
%
\title{A GALEX based search for the sparse young stellar 
population in  the Taurus-Aurigae star forming region}

\shorttitle{UV survey of Taurus}
\shortauthors{Gomez de Castro et al.}

\author{Ana I. G\'omez de Castro\altaffilmark{1}, Javier Lopez-Santiago\altaffilmark{1},
Fatima L\'opez-Mart\'{\i}nez\altaffilmark{1}, N\'estor S\'anchez\altaffilmark{1}, Paola Sestito}
\affil{AEGORA, Universidad Complutense de Madrid, Plaza de Ciencias 3,
28040 Madrid, Spain }
\and
\author{Elisa de Castro\altaffilmark{2}, Manuel Cornide\altaffilmark{2}}
\affil{Fac. de CC. F\'{\i}sicas, Universidad Complutense de Madrid, Plaza de Ciencias 1,
28040 Madrid, Spain }
\and
\author{Javier Ya\~nez Gestoso\altaffilmark{1}}
\affil{AEGORA Research Group, Universidad Complutense de Madrid, Plaza de Ciencias 3,
28040 Madrid, Spain }

\altaffiltext{1}{Fac. de CC. Matem\'aticas, Universidad Complutense de Madrid, Plaza de Ciencias 3,
28040 Madrid, Spain }
\altaffiltext{2}{Fac. de CC. F\'{\i}sicas, Universidad Complutense de Madrid, Plaza de Ciencias 1,
28040 Madrid, Spain }

\begin{abstract}

In this work, we identify 63 bona fide new candidates to T Tauri stars (TTSs) in the Taurus-Auriga region 
using as baseline its ultraviolet excess. The initial data set has been defined from the 
GALEX all sky survey (AIS). The GALEX satellite obtained images in the near ultraviolet (NUV) and far ultraviolet (FUV) bands where the TTSs show a prominent excess, compared with main sequence or giants stars. 
GALEX AIS surveyed the Taurus-Auriga molecular complex, as well as,  a fraction of the California Nebula and the Perseus  complex; bright sources and the dark clouds themselves are avoided. 

The properties of the TTSs in the ultraviolet (GALEX), optical (UCAC4) and infrared (2MASS) have been
defined using as qualification sample the TTSs observed with the  International Ultraviolet Explorer. 
The candidates have been identified by means of a mixed ultraviolet-optical-infrared excess set 
of colors; it is found that the color-color diagram FUV-NUV versus J-K is ideally suited for this purpose. 
From an initial sample of 163,313 bona-fide NUV sources, a final list with 63 new candidates to TTSs in the region has been produced. The search procedure has been validated by its ability to detect all known TTSs in the area surveyed: 31 TTSs. 

Also,  it is shown that the weak-lined TTSs are located in a well defined stripe in the FUV-NUV versus J-K diagram.
Moreover, we provide in this work a list of TTSs  photometric standards for future GALEX-based studies of the 
young stellar population in star forming regions.

\end{abstract}

\keywords{Star formation - low mass stars - pre-main sequence stars - stellar populations}

\section{Introduction}\label{intro}

T Tauri  Stars (TTSs) are precursors to Solar-like stars and planetary systems. As such, their study is fundamental for the comprehension of Solar System formation, evolution and planetary building up. Around the Earth location in the Galaxy, there is a ring of star forming regions which constitutes the main laboratory for the study of the early phases of Solar System formation. Regions like Taurus, Aurigae, Lupus, Chamaleon or Ophiuchus are within a radius of 140 pc around the Earth, in a clean area of the Galaxy, where extinction is low.  Moreover, they are at low galactic latitudes ($-20^o < b_{gal} < 20^o$) minimising the pollution of UV surveys by background galaxies. These regions form mainly Solar-like stars (though B stars are observed in Ophiuchus or A stars in Aurigae). All together, they cover a large area in the sky, being the Taurus-Auriga complex the most extended, roughly 20x20 square degrees in the sky. 

The identification of TTSs in these regions has proceeded at a fast rate. From the first observations of infrared excess and magnetically active stars in the neighbourhood of the molecular clouds (see i.e. Strom et al. 1976, Herbig \& Bell, 1989, Lada 1983) to the late use of the  large scale surveys of the Galaxy (Neuhauser et al. 1995, Wichmann et al. 1996, Li \& Hu, 1998, Kohoutek \& Wehmeyer 1999  
Rebull et al. 2011, Esplin et al. 2014).  In all these works, lists of candidates have been compiled. To search for candidates, these investigations resourced to some of the TTSs basic properties such as their proximity to the molecular clouds, the equivalent width of the H$\alpha$ line, their proper motions and kinematics  or their enhanced X-ray flux.  Final source identification resources usually to the optical to infrared spectral energy distribution (SED) for the CTTSs; the accretion disk imprint is clearly apparent in the SED. However for WTTSs spectroscopic follow-up observations are adopted as the most reliable mean for source identification; in particular, the strength of Li~I absorption is {\it the} reliable tracer of youth  (see i.e. Wichmann et al., 1996). 

The search for candidates requires to understand the distinct nature of the TTSs with respect to other galactic sources.
TTSs display excess emission in various wavelength ranges compared with their main sequence analogues. This excess is caused by the release of gravitational energy during the star formation process. The conservation of angular momentum during the gravitational collapse drives to the formation of accretion disks that channel the infall of matter onto the star. Disks get heated by the dissipation of the angular momentum excess into the small scales  through the action of the magneto rotational instability, according to the current paradigma. The magnetically  mediated interaction between the slowly rotating pre-main sequence (PMS) star and the disk drives to the production of jets and outflows (see G\'omez de Castro 2013a, for a recent review). As a result, TTSs display:
\begin{itemize}
\item an infrared to millimeter wavelengths excess caused either directly by the disk radiative output itself or through  the disk reprocessing of the stellar and jet radiation (see, for instance the first works as Mendoza 1968 or the recent works as Furlan et al. 2011 or Jensen et al. 2009)
\item an optical excess caused by the veiling continuum produced by plasma at a nominal temperature of about 7,000-10,000~K that represents the low energy tail of the ultraviolet (UV) spectrum (see, for instance, the first works as Basri \& Batalha  1990 or the recent works by Ingleby et al. 2013).
\item an ultraviolet (UV) excess produced by plasmas with electron temperatures ranging from 10,000~K to 50,000~K that shows both as a warm continuum and a distinct set of spectral indicators; excess emission is observed in neutral species (H~I, O~I, C~I), singly ionised species (C~II, Si~II, Fe~II, Mg~II, O~II) all the way to highly ionised species such as C~IV, N~V or O~VI, not to mention the H$_2$ molecular emission in the Werner bands. This excess is produced by the sheared magnetosphere-disk boundary layer and the magnetospheric accretion flow with a non-negligible contribution from the outflow and the inner disk (see
G\'omez de Castro 2009 for a detailed review on the UV spectrum of the TTSs).
\item An X-ray excess (see, for instance, from the first works by Montemerle et al. 1983 to reviews such as Preibish 2004) 
produced by the strong magnetospheric activity,   the expanded corona and the accretion shocks
at the point of impact, where the gravitational energy of the infalling material is finally released into heating ({\it i.e.} Lamzin 1998, Gullbring \& Calvet 1999).
\end{itemize}

In spite of the relevance of the UV excess of the TTSs compared to that of their main sequence analogues, most of the searches for candidates have been run at either X-ray (see, for instance, Feigelson et al. 2005, Neuhauser et al. 1995) or infrared wavelengths (Furlan et al. 2011, Takita et al. 2010, Padget et al. 2006). Only the southern area of the  Taurus complex was surveyed to search 
for TTSs in the UV (Findeisen \& Hillenbrand 2010; hereafter FH).  The release of the GALEX All Sky Survey (hereafter GALEX/AIS) has provided an amazing wealth of information on the UV characteristics of the galactic stellar population (Bianchi et al. 2014), in general, and of the Gould's Belt or the local supercomplex of star formation (G\'omez de Castro et al. 2011), in particular.   Among all the possible regions of star formation, the Taurus-Auriga complex is the prototype of low mass star formation and the best studied (Kenyon et al. 2008). The complex contains many dark nebulae, Herbig-Haro jets and outflows, and young stars. With few massive O or B stars, the clouds are an excellent place to study the formation of Sun-like stars, in contrast with the Orion Nebula Cluster, at 450~pc, where a significant population of young massive stars resides.

The GALEX/AIS survey of the Taurus-Auriga molecular complex consists of  197  GALEX fields 
covering  380 square degrees in the sky; more than 3 times larger than the FH survey. About 163,000 NUV sources have been detected allowing detailed studies of the stellar population in this area of the Galaxy, as well as, searching for young stars. The area mapped by GALEX also includes the  northern tip (in galactic coordinates) of the Hyades stellar association.

In this work, we have developed a set of new criteria, different than those proposed by FH's, to search for TTSs in UV surveys. 
Our method makes use of the known spectral energy distribution of the TTSs in the UV  to derive,  the UV magnitudes of the TTSs in the GALEX bands.  Once the method is validated, we have used a combination of UV and IR colors to discriminate the TTSs from the background stellar population. 

Our working sample consists of 163,313 UV sources that have been  cross-correlated with  optical and IR catalogues to produce a number of color-color and color-magnitude diagrams. From that, a total of 63 new candidates to TTSs have been found.  In Sect.~2,  the photometric properties of the TTSs in the GALEX bands are calculated. In Sect.~3, the main characteristics of the GALEX survey are outlined; the multi-band photometric criteria are defined and applied to the GALEX sample to produce a preliminary list of candidates. In the process, 31 known TTSs within the area mapped by the GALEX-AIS are retrieved. As a result, a more robust data base is established to qualify the TTSs candidates in the GALEX~AIS survey. This data base is used in Sect~4 to further refine the search and produce a final list of 63 new candidates, most of them class II-III sources (Wilking \& Lada 1983). A short discussion on the sources location and a brief summary are provided in Sect.~5 and 6, respectively.

\section{Properties of the TTSs  in the GALEX bands}

GALEX was a 50-cm primary space telescope with a Ritchey-Chretien mounting feeding simultaneously two detectors sensitive to the near and far  UV  making use of a multilayer dichroic beamsplitter. The near UV or NUV channel works in the 1770-2730~\AA\ spectral range and the far UV or FUV channel in the 1350-1780~\AA\ range; the transmittance curves of the channels are shown in Fig.~1, together with the UV spectrum of
T~Tau, the prototype of TTS. The characteristics of the plasma sampled by the NUV and FUV bands are quite distinct:

\begin{itemize}
\item The spectrum of the TTSs in the FUV band contains mainly  emission lines covering a very broad range of plasma temperatures; from the molecular bands of H$_2$ (Herczeg et al. 2002, France et al. 2011) to the resonance transitions of Si~IV, C~IV, C~I and He~II (G\'omez de Castro \& Franqueira 1997, Valenti et al 2000, Yang et al 2012,  G\'omez de Castro \& Marcos-Arenal 2012, Ardila et al. 2013, G\'omez de Castro 2013). The peak of the band is at 1500~\AA\ hence, 
GALEX FUV band radiation is dominated by the fluorescence H$_2$ radiation and the C~IV, He~II and Si~IV features. Radiation in these lines correlates with accretion rate (see, France et al. 2012, Ardila et al. 2013 and G\'omez de Castro 2013b, for the most recent references in the field). In cool stars, radiation in the GALEX FUV band correlates with the stellar activity, as measured by the H,K Ca~II index (Findeisen et al., 2011) 

\item The spectrum of the TTSs in the NUV band contains mainly radiation from warm plasma at temperatures between 3000~K and some 10,000~K; the dominant features being the Mg~II[uv1] multiplet,  the Fe~II resonance multiplets [uv1], [uv2] and [uv3], the C~II] feature at 2326~\AA\  and the short wavelength tail of the Balmer continuum. The transmittance curve of the band enhances the radiation from the Fe~II multiplets and the 2326~\AA\ feature with respect to the Mg~II radiation, the strongest line in the NUV spectrum. For instance, in T~Tau, 38\% of the NUV radiation is in the C~II]$_{2323}$, Mg~II and Fe~II lines. Both line radiation and continuum correlate with accretion rate and chromospheric activity.
\end{itemize}

These trends are readily shown in Figs.~2 and 3. In Fig.~2, the C~IV line excess radiation in the TTSs is compared with that of main sequence cool stars, including the very active M-dwarfs (from G\'omez de Castro \& Marcos-Arenal 2012). As shown TTSs are several orders of magnitude brighter, allowing to detect them at the 140 pc distance of the  TMC. Therefore, radiation in the FUV band can be used as reliably as H$\alpha$  or X-ray radiation surveys to identify the TTSs by their unusual strength. In Fig.~3, the M$_V$ versus UV-V color is plotted for simultaneous near UV and optical observations obtained with the IUE of TTSs  (G\'omez de Castro 1997). The location of the late type (K3 or later) TTSs in the diagram is plotted against the location of  cool main sequence stars.
A quick inspection shows that the UV-V color decreases as the TTSs approach the main sequence but that the color 
excess is negligible in the Weak-lined TTSs (WTTSs).

To derive the expected photometric properties of the TTSs in the GALEX bands, we have used the sample of the TTSs observed with the IUE in low dispersion mode and compiled 
by G\'omez de Castro \& Franqueira 1997 for the IUE Uniform Low Dispersion Archive Guides (hereafter GdCF); only the 21 TTSs with high SNR spectra over the whole range have been selected for this purpose (see Table~1). 
Their IUE spectra have been multiplied by the normalised GALEX transmittance function in the  FUV and NUV windows to compute their FUV and NUV AB magnitudes using Morrisey et al. (2007) conversion:
\noindent
$$
NUV = -2.5 \times \log \bigg( \frac{\rm Flux NUV}{2.06  \times 10^{-16} {\rm erg\,s}^{-1}{\rm cm}^{-2} {\rm \AA}^{-1} } \bigg) + 20.08
$$ \\
\noindent
$$
FUV = -2.5 \times \log \bigg( \frac{\rm Flux FUV}{1.40 \times 10^{-15} 
{\rm erg\,s}^{-1}{\rm cm}^{-2} {\rm \AA}^{-1}} \bigg) + 18.82
$$

This IUE sample is scarce in terms of WTTSs and M-type CTTSs due to the sensitivity of the IUE. Also note that given the spectral energy distribution of the TTSs, their FUV magnitude is rather low (especially for WTTSs), as a result it is challenging to detect cool WTTSs in the GALEX FUV band (see below).

\section{The GALEX survey of the TMC}

The baseline of the GALEX All Sky Survey was completed in 2007.  GALEX AIS covers 26,000~deg$^2$ ($\sim 63$\% of the Sky) and provides broadband imaging in two UV bands. GALEX obtained 197 images on the Taurus molecular cloud for a total coverage of $\sim 200$ deg$^2$ (see Fig.~4); note that the GALEX field of view is circular with radius 0.6$^o$.
The GALEX mission provides as output products for each tile (or image in the AIS survey) several files containing the pipeline processed images in the FUV and NUV bands, the intermediate calibration files and the catalogue of sources identified in each pointing; the sources in the catalogue are identified with the SExtractor procedure (Morrissey et al. 2007) and their FUV and NUV magnitudes are provided. Typically, the number of sources in the catalogue outnumbers by a factor of 2-3, the number of sources that can be identified as such, from a simple inspection of the images. The detection procedure clearly suffers from overestimation of actual sources. The number of spurious detections is very large. 

To overcome this problem we have designed a strategy that guarantees no TTSs are lost in the cross-identification procedure. TTSs are stars with photospheric spectral types F to M and an infrared excess that decreases as stars approach the main sequence. As an example, we display in Fig.~5, the spectral energy distribution of AK~Sco  located at 145~pc from the Sun.
The prominent FUV and infrared excesses are readily recognized over the stellar atmosphere radiative output predicted by the Kurucz model of an F5 star (Alencar et al. 2003). Thus, a cross-identification with galactic sources from the Fourth USNO CCD Astrograph Catalog - U.S. Naval Observatory (UCAC4; Zacharias et al. 2013) and the 2MASS surveys (Strutskie et al. 2006) is a reliable mean to remove spurious sources without loosing TTSs in the procedure. 

{\it Bona fide} GALEX sources have been identified by having a 2MASS counterpart within a search radius of 3~arcsec. This search radius was carefully selected after a precise study of the shift between 2MASS and GALEX sources in the TMC (see G\'omez de Castro et al 2011, for details). 
After cross-match all sources satisfied  $NUV \lesssim 22 $~mag  what corresponds to the $\sim 3\sigma$ detection threshold for a typical exposure time of 100~s
(Bianchi et al. 2013). Magnitude completeness limits for 2MASS are 15.8, 15.1 and 14.3 mag in J, H and K$_\mathrm{s}$
bands, respectively. These limits are well above the magnitude of M-type stars at the distance of the Taurus-Auriga molecular complex. The J absolute magnitude for a 4~Myr old M5 star is $\sim 5.5$ (e.g. Siess et al. 2000). At a distance of 140 pc, the magnitude of such star $J \sim 11.5$~mag. Even in the case of notable ISM extinction 
$A_\mathrm{J} = 4$~mag, the star would be detected by 2MASS at the distance of the molecular cloud. 
Similarly, UCAC4 is complete from the brightest stars to about $R = 16$~mag. This means that 
UCAC4 is complete for mid-M stars at the distance of the Taurus cloud for extinctions
$A_\mathrm{V} < 2$~mag. As we do not expect large extinctions for our stars, because they were 
selected from the UV, we can conclude that we are complete for members of the star-forming 
region for the entire luminosity function, from the more massive to the less massive stars. 
This does not assure completeness  down to M dwarfs at 140 pc (e.g. Lepine \& Gaidos 2011).  

The GALEX AIS survey is however, deep enough to reach the TTSs population in the TMC. 
Bright CTTSs such as RW~Aur, K3 spectral type with $A_V = 0.5$~mag (Ingleby et al. 2013), is found to have $FUV = 15.9$ and $NUV=13.8$  in the GALEX bands and a WTTS, such as V836~Tau (K7) with $A_V = 1.5$ (Ingleby et al. 2013), has  $FUV = 20.8$ and $NUV=19.9$. Both stars are at 140~pc from the Sun.
As shown below, the GALEX AIS limiting magnitude in the, very demanding, FUV channel is 22.3 for this survey. In summary,
the survey must be complete to the TMC distance for CTTSs and most WTTSs, except for embedded sources and very late types. Non extincted sources, as bright as RW~Aur could be detected as far as 2.2~kpc. Otherwise, RW~Aur would be detectable even if extincted by  $A_V = 2.7$~mag instead of 0.5~mag; the Fitzpatrick (1999) extinction law has been assumed (see Yuan et al. (2013) for a discussion about possible deviations of the ISM extinction law in the NUV and FUV bands).

After the cross-correlation, we kept a total of 163,313 UV sources as reliable detections. All them have, at least, a 2MASS counterpart at less than 3 arcsec. Fig.~6 shows the density map of our sources in the selected region. The 2MASS extinction map by Lombardi et al. (2010) is overplotted. Spatial binning is 3~arcmin$^2$ and the color code ranges from 0 to 8 sources per arcmin. The stellar density is clearly not homogeneous. Many regions suffer from large extinction, coinciding with the location of dense molecular clouds. However, the density of UV stars is also low in some regions  where extinction is low as inferred from CO or 2MASS maps. We refer the reader to G\'omez de Castro et al. (2015) for a detailed discussion on this issue. Notice that the GALEX survey is significantly shallower than the 2MASS survey for galactic sources because extinction prevents UV radiation to propagate long distances in the galactic plane. Henceforth, no UV candidates to TTSs are lost in the cross-matching procedure.

\subsection{Basic selection criteria for TTSs}

Only 1\% , from a grand total of 163,313 sources, are detected both in the NUV and FUV bands; they constitute our working sample since our goal is to search for {\it " bona fide" candidates}  instead of  guaranteeing that no TTS in the field is lost in the search. In practice, this means that some weak TTSs or brown dwarfs may not be included in the list, being too faint to be detected in the FUV band. However, even with this restrictive criterium, the known TTSs in the area mapped by the GALEX~AIS are retrieved (see below) some of them, like LkCa~1 lying very close to the sensitivity limit in the FUV band.

The (FUV-J, J-K) and (NUV-J, J-K) diagrams are plotted in Fig.~7. Notice that the colors of the vast majority of the sources agree with the predictions of the Kurucz models for main sequence stars (compare Fig.~7 with Fig.~4 in FH). The location of the IUE qualification sample is clearly different from that of the dominant galactic population in the plot;  the J-K color of the classical TTSs  exceeds that of the field (main) sequence stars by more than $5\sigma$ according to FH's Fig.~4. 

At this point, it is worth remarking the differences between the procedures used in FH and in this work to identify TTSs candidates. In this work, we define candidates by {\it “TTS class membership”} while FH define their TTSs candidates by not-belonging to the {\it “main-sequence class membership”}, this last one being defined from a sample of field stars. FH used field stars photometry and an iterative procedure to define the locus of the main sequence stars by a linear regression line in the (FUV-J) and (NUV-J) versus (J-K) diagram. However, we identify as candidates those sources that {\it share the same color than the well-known TTSs of the IUE template sample}; thus no further assumptions are introduced. According to our criterium, TTSs candidates have to satisfy 7~mag$ < FUV-J < $12~mag  and  4~mag $<  NUV-J < 9$~mag however, some  FH's candidates lie outside these boundaries. The extinction arrows have been plotted using 
the coefficients derived by Yuan et al. (2013) for the Fizpatrick extinction law. Notice that the TTSs in the qualification sample have well studied extinctions with $A_V = 0.1 - 1.5$ mag; the two WTTSs being very close to the field stars location.

At this point it is worth reminding the main sources of extinction in the TTSs environment: the circumstellar (CS) and the interstellar medium (ISM). For the TTSs in the TMC, ISM extinction is mainly caused by the molecular cloud; the 
complex is located at the outskirts of a low density bubble in the local ISM (see {\it i.e.} the Ungerechts \& Thaddeus 1987 CO survey of the area). At early stages, TTSs can be deeply embedded in the cloud and totally occulted at UV or optical wavelengths; as indicated above even a very bright TTSs, such as RW~Aur,  would not be detected in the 
FUV band for $A_V=2.7$~mag. For this reason, the GALEX survey has avoided the dark molecular cores and only covers the halo where extinctions are significantly low, $N_H < 8.8 \times 10^{20}$~cm$^{-2}$ or $A_V < 0.5$~mag, as inferred by comparing Fig.~5 with Lombardi et al 2010 (see also Gomez de Castro et al. 2015). CS extinction is linked to the
nature of the TTSs (dusty disk atmospheres, orientation effects, see {\it i.e.}  Watson et al. 2007 or Dullemond et al. 2007) and, as such, it is used to qualify the candidates.  

The generic feature characterising the TTSs is the co-existence of  FUV, NUV and infrared excess; this is also shown in Fig.~8 where the optical B, V, R bands are used as fiducial, pivot bands for the photospheric radiation.  Optical photometry has been retrieved after cross-correlation with UCAC4 for the candidate sources (see above). 

The separation between the TTSs and the galactic stellar population is clean unless for weak lined TTSs (WTTSs). Sources like AB~Dor or HD~283572 (see Table~1) are located in the same region than main sequence stars in the color-color diagrams involving the NUV band. 

In Fig.~9, UV versus infrared colors are directly plotted. The UV color, FUV-NUV, is a mixed index that includes: [1] the excess of the hard UV radiation (C~IV, Si~IV, He~II) over the soft UV radiation (Balmer continuum, Fe~II, C~II, Mg~II), as well as, [2] the excess radiation produced by the reprocessing of Lyman~$\alpha$ photons in the disk (see i.e. France et al. 2012)
with respect to the photospheric and chromospheric radiation flux. This index is plotted against the J-K and the H-K infrared colors. H-K is considered to be a good estimator of the accretion rate (see i.e. Meyer et al. 1997 for details) however, WTTSs are not well discriminated 
from main sequence stars in H-K. The J-K color is a much more reliable tracer (see Fig.~9). Therefore, we will consider the
FUV-NUV vs J-K diagram our prime diagram to select TTSs candidates from the GALEX survey. 
Notice that the impact of interstellar extinction in the candidates selection is small in the color-color diagrams. Also,
$A_V > 2$~mag to shift a given source from the field stars stripe  into the TTSs locus within the diagram (see Sect.~3.3 for a discussion
on WTTSs). 

We have also considered other diagrams such as NUV-R vs J-K and NUV-H vs J-K to search for candidates. They are useful to characterise the young stellar population in farther regions (such as Orion) where most of the TTSs are not detectable in the FUV band.
A preliminary list of 315 TTSs candidates has been generated including all sources that satisfy at least one of the following criteria: 

\begin{itemize}

\item {\it Criterium I: FUV-NUV vs J-K}. TTSs satisfy $0.5 < J-K < 2.4$ and $0.4 < FUV-NUV < 4.6$ . 

\item {\it Criterium II: FUV-NUV vs H-K}. TTSs satisfy $0.2< H-K < 1.2$ and  $0.4 < FUV-NUV < 4.6$ .  

\item {\it Criterium III: NUV-R vs J-K}. TTSs satisfy $0.5 < J-K < 2.4$ and $1.5 < NUV-R < 8.2$ . 

\item {\it Criterium IV: NUV-H vs J-K}: TTSs satisfy $0.5< J-K < 2.4$ and  $4.2 < NUV-H < 11.0$.  

\end{itemize}

\noindent

This list has been purged from spurious sources using the services of the  Stellar Data Base at the Centre de Donn\'ees Stellaires in Strassbourgh (CDS). CDS provides information on the object type\footnote{\url{simbad.u-strasbg.fr/simbad/sim-display?data=otypes}} that it is defined as a hierarchical classification, which emphasises the physical nature of the object rather than a peculiar emission in some region of the electromagnetic spectrum or the location in peculiar clusters or external galaxies. The two main channels for feeding the database are the daily scanning of papers published in the astronomical literature providing new references and new identifiers for existing objects, as well as the complete (or partial) folding of selected catalogues into the database serves as a basis for improving the completeness and multi-wavelength coverage of the database (see Wenger et al. 2000). Only objects classified as TT*, Em* or without CDS identification are kept. The CDS bibliographic service has also been used to refine further the identification.  Most of the stars classified as Em* are found to be WTTSs. Also some TTS candidates from the WISE survey (Rebull et al 2011) or from old  ROSAT-based surveys  (Wichmann et al. 1996, Li \& Hu 1998) are recovered in our GALEX-based search for candidates. 

The selection procedure is robust; all known CTTSs included in the fields mapped by the GALEX AIS  are recovered in this list of candidates.
This includes the CTTSs:   HBC425, CI~Tau, GH~Tau,V807~Tau, ZZ~Tau, DK~Tau, IQ~Tau, FP~Tau, CX~Tau,
CW~Tau, FM~Tau, Elias~12, V773~Tau, V836~Tau, IRAS04108$+2910$, GM~Aur, RW~Aur and UY~Aur.  

Both GM~Aur and RW~Aur were also part of the IUE qualification sample of TTSs.  The discrepancy  between their GALEX FUV and NUV magnitudes and those  calculated from the IUE spectra  
is smaller than the typical variability range of the CTTSs: $< 0.75$~mag or a factor of 2  in flux. The strongest discrepancy is found in the FUV magnitude,  
$\Delta FUV = FUV_{\rm GALEX} - FUV_{\rm IUE} = 0.5$~mag for RW~Aur but this can be easily accounted by the large intrinsic variability of this source 
(see e.g. G\'omez de Castro \&  Verdugo 2003). 

Well-known WTTSs are  also retrieved with our search criteria namely, HD30171, RXJ0437.4$+$1851B, V1207~Tau, V1195~Tau, LkCa~1, HD~281691, LkCa~19, V600~Aur and V583~Aur.  
Our candidate list includes some WTTSs discovered recently from the analysis of the Spitzer data (IRAC) of the TTSs candidates identified in the WISE survey (Rebull et al. 2011), 
{\it i.e.} the binary BS~Tau or 2MASSJ043601+1726120 (Esplin et al 2014).   They however, only satisfy our prime criterium (criterium I).

In summary, 31 out of the 315 candidates in the preliminary list are well known TTSs. This extends the IUE working sample, especially in terms of WTTSs, permitting to analyse the overall photometric properties of the TTSs as a class from the FUV to the IR (see Sect.~4).

\subsection{UV color-magnitude diagrams}

UV color-magnitude diagrams are represented for the candidates in Fig.~10 and 11.
The depth of the GALEX survey is best visualised in Figure~10; the limiting FUV magnitude of 22.3 mag shows as the limiting boundary $NUV < 22.3 - [FUV-NUV] $. 
The TTSs from the IUE sample have also been plotted; their NUV and FUV magnitudes have been corrected to the 
TMC distance (140~pc) for the figure.  Extinction arrows are plotted for a 7,000~K spectrum (as in
Yuan et al. 2013 for the Fitzpatrick extinction law) and for T Tau; both for R=3.1. Note the sensitivity of the 
color correction to the underlying energy distribution. 

Let us examine firstly, the NUV vs. FUV-NUV diagram. The WTTSs are close to the 
sensitivity boundary, they are the reddest sources in the plot.
Also note that the CTTSs are not only brighter but also significantly 
bluer than the WTTSs. We have computed statistic distribution function to this boundary
for CTTSs and WTTSs to quantify this qualitative behaviour.  The WTTSs distribution
has a single peak with $d_{lim} = 1.31 \pm 0.65$~mag however, the CTTSs have a 
a main peak at $d_{lim} = 3.15$~mag and a broad tail to lower distances. This broad distribution 
is associated with the CTTSs  spectral type, being  M-type CTTSs located closer to the
limiting magnitude boundary than earlier types. 

In the FUV vs. FUV-NUV plot (Fig.~11), there is not such a clear trend. 
Most of the  candidates are close to the limiting magnitude boundary. It is worth noticing that
stars as active as AB~Dor should be detectable at the distance of the TMC from the GALEX~AIS. 

To get further insight into the low luminosity limit, we have retrieved from the HST archive, the ultraviolet spectrum of the only brown dwarf observed, 2MASS~J12073346~39322539\footnote{The data set used was lb4p01010 and the observations were obtained with the Cosmic Origins Spectrograph and grating G140L in December 3rd, 2009 (see France et al 2010).}, hereafter 2M1207.  This is an M8 brown dwarf located in the TW~Hya association at 52.4~pc (Webb et al 1999). Unfortunately, the observation only covers the FUV Galex band. GALEX~AIS limiting magnitude in this band, 22.3, corresponds to a flux of $5.74 \times 10^{-17}$~erg~s$^{-1}$~cm$^{-2}$~\AA $^{-1}$ which is slighty higher than the integrated 2M1207 flux in the FUV band ($4.01 \times 10^{-17}$~erg~s$^{-1}$~cm$^{-2}$~\AA $^{-1}$); 2M1207 FUV magnitude is 22.7. In summary, brown dwarfs similar to 2M1207 are very close to the GALEX~AIS detection limit, if located at a distance comparable to the mean distance of the Hyades open cluster, 45~pc, (de Bruijne et al. 2001), as shown in Figure~11.

Finally, note that the separation between CTTSs and WTTSs in Fig.~10 is nearly vertical, as the  extinction law is. This shows that there is an  {\it extinction-accretion degeneracy}; 
extinction makes the sources weaker  while accretion produces an increase in the TTSs luminosity both, with a mild signature in the FUV-NUV color. 

\subsection{Refining the sample: hot stars removal}

Two, out of the four criteria indicated in Sect.~3.1, do not impose any constraint to the FUV-NUV color.
As a result, the first list of candidates includes some hot sources with colors $FUV-NUV < 0.4$. Some 
of these could be just hot stars in the field. In principle, we could remove them directly from the list of candidates
but we have preferred to use a more sophisticated procedure that takes into account the SED (and not just one color). 
However, we have detected some well known TTSs, such as HBC~425 (with $FUV-NUV > 0.4$) that would be missed
with this procedure. Thus, we have decided to the SED to make a simple determination of the effective temperature
and remove from our list of candidates sources with $T_e > 7,500$.

We have built the optical SED for the whole set of candidates using UCAC4 photometry; the optical range is the 
best suited to estimate the stellar effective temperature for spectral types F-M.
After, we have used the VOSA tool of the Virtual Observatory\footnote{svo2.cab.inta-csic.es/theory/vosa/}  to fit 
photospheric models to each  candidate and determine its effective temperature.

VOSA combines the multiwavelength data from generic archives and catalogues within a  VO environment
to find the best fit to a predetermined set of theoretical models. The provided “best” fitting model is 
the one that minimises the value of the reduced $\chi ^2$ defined as:
$$
\chi ^2 = \frac{1}{N-P} \sum \bigg( \frac {1}{\sigma _0 ^2} (Y_0 - M_dY_m)^2 \bigg)
$$
where $Y_0$ is the observed flux, $\sigma _0$ the observational error in the flux, 
$N$ the number of photometric data points, $P$ the number of parameters being fitted, $Y_m$ the
theoretical flux predicted by the model and $M_d$ the multiplicative dilution factor, defined as 
$(R/D)^2$ (for models from Hauschildt et al. 1999, Allard et al. 2011 and Chabrier et al. 2000),
$R$ being the radius of the source and $D$ the distance to the object (we refer the reader
to Bayo et al. 2008 for a more detailed description of the method and the goodness of the estimates).
Stellar models implemented in VOSA are Kurucz, NextGen or DUSTY00 stellar models.
The DUSTY00 synthetic models account properly for the effect of dust 
grains formation in the atmosphere of very cool stars and brown dwarfs. NextGen models represent 
a later elaboration in the synthetic spectra of cool stars (see Allard et al. 2012, for further
details).  

As standard output VOSA produces several parameters including effective temperature.
Taking into account the few data points available per star, we have worked with a reduced set of 
free parameters fixing metallicity to solar and extinction to 0~mag. Note that extinction causes 
the SED to become redder and thus cooler. As a result, reddened hot stars may be kept in the sample 
but certainly those sources with high $T$ are not TTSs since the UV excess of the TTSs is not 
high enough to mimic early spectral types (see Fig.~5).

After rejection of sources with effective temperatures above 7500~K, our list of candidates has reduced 
to 212 sources.

\section{Photometric properties of the TTSs from FUV to IR: further refinement of the TTSs candidates list}

As pointed out in Sect.~3, the detection of known TTSs in the TMC has increased significantly the photometric 
data base of TTSs in the GALEX bands from our primitive IUE sample, especially concerning WTTSs. 
The new sample includes 15 WTTSs and 31 CTTSs,  10 of them M-type stars providing a better statistics for the candidates search. 
As shown in Fig.~12, the best diagram to discriminate TTSs from galactic field stellar population keeps being the FUV-NUV versus J-K diagram, however new trends can now be identified such as,
\begin{itemize}
\item the locus of the WTTSs is well defined by the line:
$FUV-NUV = (-3.88 \pm 0.61) (J-K) + (5.64 \pm 0.55)$, with RMS=0.59. The dispersion of the WTTSs 
around this linear regression line has been computed and used to define the $3\sigma$ boundary
marked with dashed lines in Figure~12.
\item the locus of the CTTSs in J-K is well defined by a normal distribution centered at  $(J-K) = 1.4$ with dispersion, $\sigma = 0.4$. However, in
FUV-NUV the only constraint set by the data is: $0 < FUV-NUV < 3.5$, very similar to our criteria in Sect.~3. 
\end{itemize}

These constraints have been used to reduce the list of 212 candidates (see Sect.~3) to just 63  (see Table~2) by basically, removing from the list of candidates those stars with small J-K and FUV-NUV colors. Notice that there is some degree of overlapping between WTTSs and CTTSs that corresponds to hot WTTSs with low infrared excess and CTTSs with small infrared excess, {\it i.e.} type II-III TTSs according to the standard infrared SED classification (Wilking \& Lada 1983).

Using the constrains summarised above, the candidates have been classified in Table~2 as:{\it candidates to WTTS}, 
{\it candidates to CTTS} or just {\it candidates} depending on their location in the FUV-NUV vs. J-K diagram. 
WTTSs candidates are stars within 3$\sigma$ of the WTTSs regression line that satisfy $FUV-NUV > 3.5$,
to ensure they are not in the area where the WTTSs and the CTTSs overlap in the diagram.
CTTSs candidates are stars satisfying  $ 0 < FUV-NUV < 3.5$ and lying outside the $3\sigma$ boundary around the WTTSs regression line. The rest of the sources have just been classified as TTS candidates.

Notice that both WTTSs and CTTSs cover a broad range in FUV-NUV. In WTTSs, the FUV-NUV excess correlates well with the J-K excess
but this is not the case for CTTSs. These stars are surrounded by accretion disks that are luminous in the infrared bands breaking the
expected correlation between the UV and infrared excess, associated with the stellar spectral type.
In an accretion dominated SED, the correlation between the UV and IR excess should have a different functional form than
observed in stars; moreover, inclination effects and extinction would affect in a different manner to the disk radiation and 
the UV radiation from the accretion shocks and the stellar magnetosphere;  CTTSs with similar accretion 
rates/luminosities and J-K color may show small or high FUV-NUV values depending on  whether the disk is seen face-on or pole-on, respectively, see {\it i.e.} the determination of the disk inclination in the so-called silhouette disks (O'Dell et al., 1993).  
In fact, the uncorrelated combination of all these effects is the most likely cause of the spreading observed in Fig.~14.

\section{Discussion}

The distribution of UV TTSs and candidates on the TMC is displayed in Fig.~13; the already known CTTSs and WTTSs 
are labeled. The new sources are concentrated in the filaments  around
$l \simeq 170^o-174^o$ and $b \simeq -4^o - -10^o$ between the Auriga and the  California complexes.  
Notice that there is not a correlation between the spatial density of UV sources and TTSs candidates (compare Fig.~5 with
Fig~15).  The proper motions of the PMS stars in the region from Doucourant et al. (2005) are 
overlayed in Fig.~15 for reference. They indicate the average direction of the TMC motion in the plane of the sky. As pointed out
by G\'omez de Castro \& Pudritz 1992, the good kinematical coupling between the TTSs and the molecular cloud in radial velocity
can be extrapolated to the filaments kinematics in the plane of the sky.

A global scenario was proposed by G\'omez de Castro \& Pudritz (1992) for star formation in Taurus. According to it, the driving energy source of the cloud hydromagnetic turbulence is the cascading of a large magnetic wave unveiled by the polarization maps of the region (Moneti et al. 1984). Recent evidence of hydromagnetic waves action in the Taurus filaments
has been found by Heyer \& Brunt (2012) from the analysis of the CO data in Davis et al. (2010). According to this
global scenario the youngest sources  should be at the cloud front where the molecular gas is concentrated.
The most evolved TTSs are thus expected, to be located at the rear of the cloud where most of the new candidates to TTSs
unveiled by this work have been found.

\section{Conclusions}

From an initial sample of about 350,000 UV sources in the TMC according to the GALEX-AIS catalogue,  a list of 63 bona fide candidates to TTSs has been 
produced. For this purpose, a first set of selection criteria have been defined based on the UV photometric properties of the TTSs to create an initial set of 315 candidates to TTSs; the SIMBAD service has been used to guarantee that only stellar unknown sources, or known sources classified as TTSs or Emission line stars are included. 
We have validated the procedure by recovering the 31 known TTSs in the area of the survey. 

As an additional result, we have generated a list containing 15 WTTSs and 31 CTTSs  that can be used as a qualification sample for UV-IR searches of TTSs in star forming regions. 
This qualification sample has been used to statistically improve the candidates selection criteria 
producing the final list of 63 bona fide TTSs candidates.

\acknowledgments
This research was partially supported by grant AYA2011-29754-C03-01 from the Government of Spain.

{\it Facilities:} \facility{GALEX}, \facility{IUE}, \facility{SIMBAD},\facility{2MASS},\facility{VO},\facility{UCAC4}, \facility{WISE} .


\newpage

\begin{figure}[tb]
\includegraphics[width=17cm]{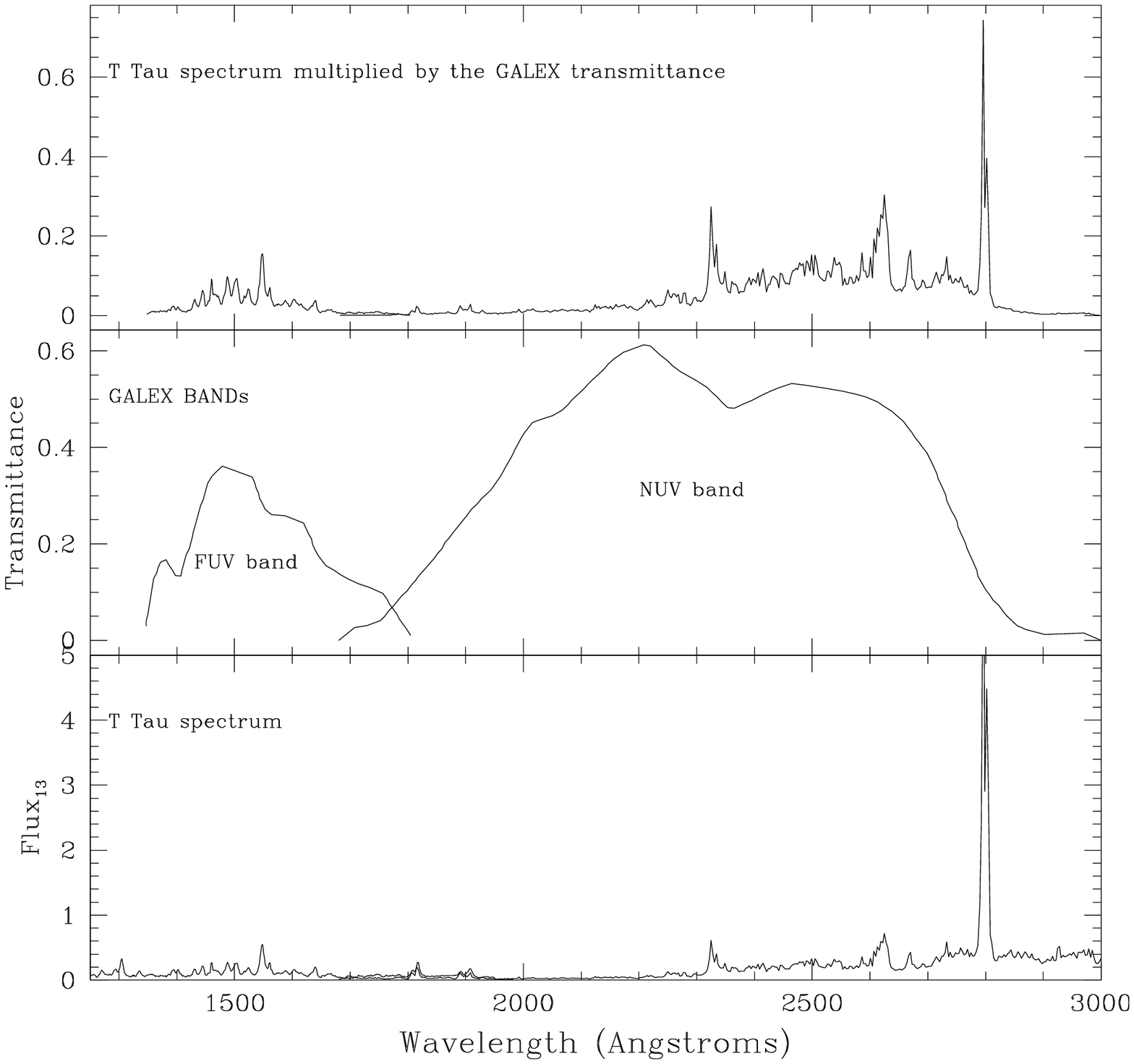}
\caption{T Tau spectrum (bottom), GALEX FUV and NUV channels spectral response (middle) and T Tau spectrum multiplied by the GALEX spectral response. The stellar flux in the bottom panel is given in units of $10^{-13}$erg~cm$^{-2}$~s$^{-1}$~\AA $^{-1}$. }
\label{GALEX-TTS}
\end{figure}

\newpage

\begin{figure}[tb]
\includegraphics[width=17cm]{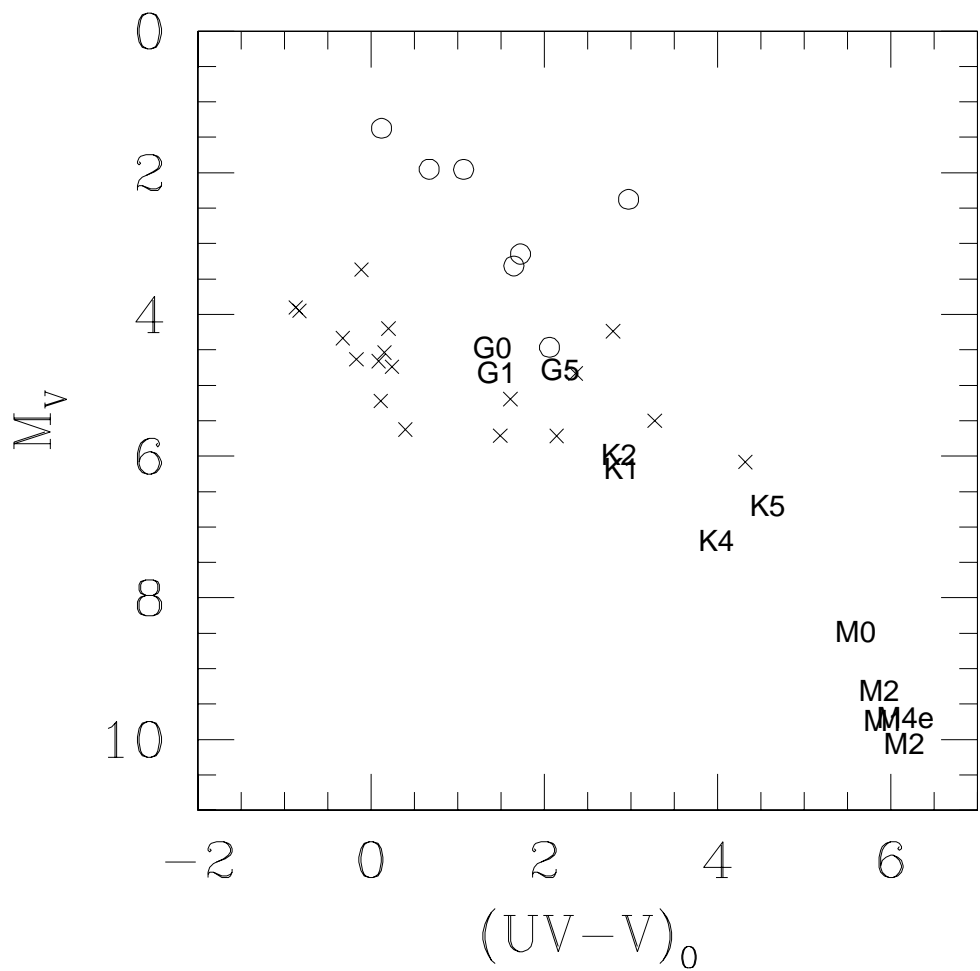}\\
\caption{Some characteristics of the UV excess in the TTSs. The normalised  fluxes are represented versus the stellar luminosities for the C~IV line (similar relations
are observed for other spectral tracers in the FUV band (see G\'omez de Castro \& Marcos-Arenal 2012).  TTSs and main sequence cool stars are plotted with triangles 
and circles, respectively. Me stars are found to be as strong in the UV as WTTSs.  Error bars for the UV normalised  fluxes of the TTSs are indicated.  }
\label{UVexcesslines}
\end{figure}

\begin{figure}[tb]
\includegraphics[width=17cm]{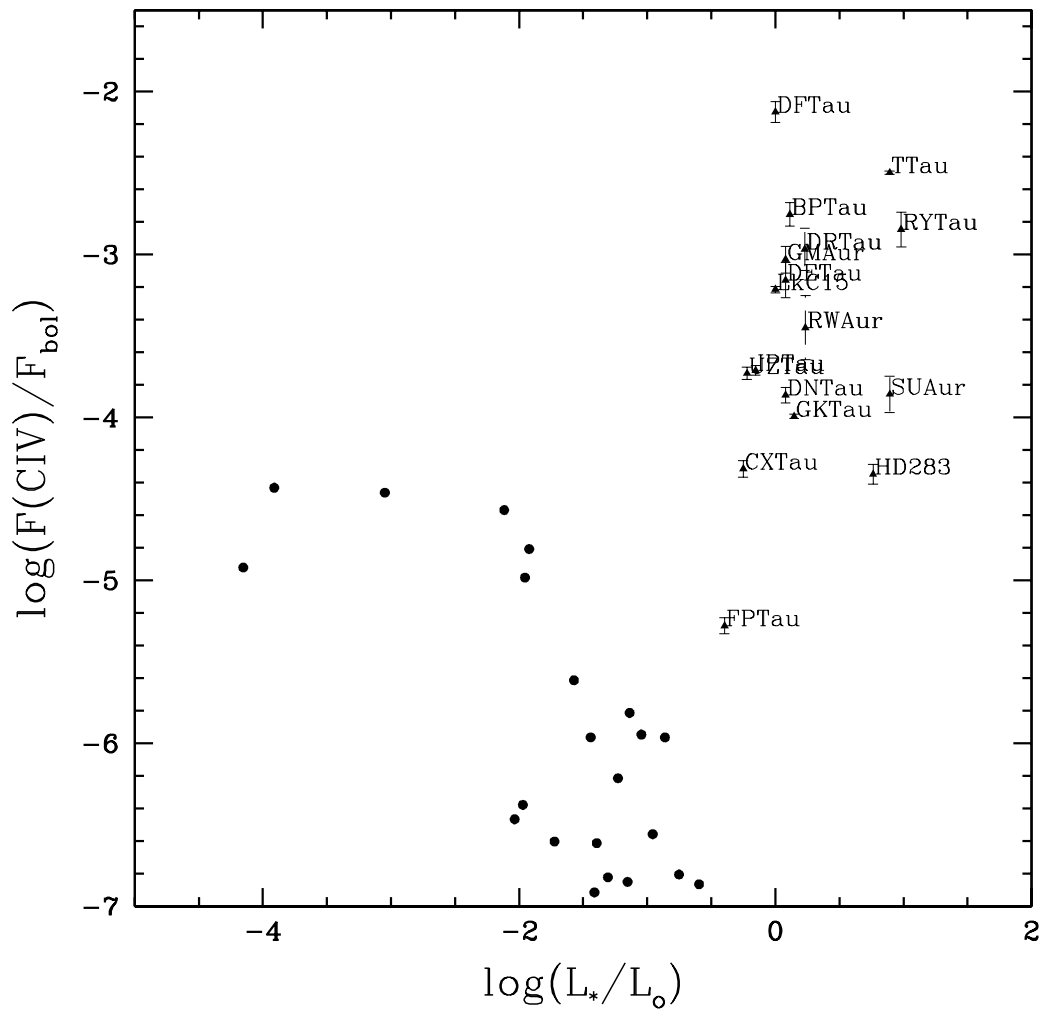}
\caption{Some characteristics of the UV excess in the TTSs. (UV-V, V) color -- magnitude diagram for the 
T Tauri stars observed with the IUE in the Taurus region. The crosses represent cool TTSs (spectral types later than $\sim $ K3) and the open circles warm TTSs (spectral types
earlier than $\sim$ K3). The location of the main sequence is marked also from IUE observations. The TTSs closer to the main sequence are the WTTSs (from G\'omez de Castro 1997).}
\label{UVexcess}
\end{figure}

\newpage

\begin{figure}[tb]
\includegraphics[width=17cm]{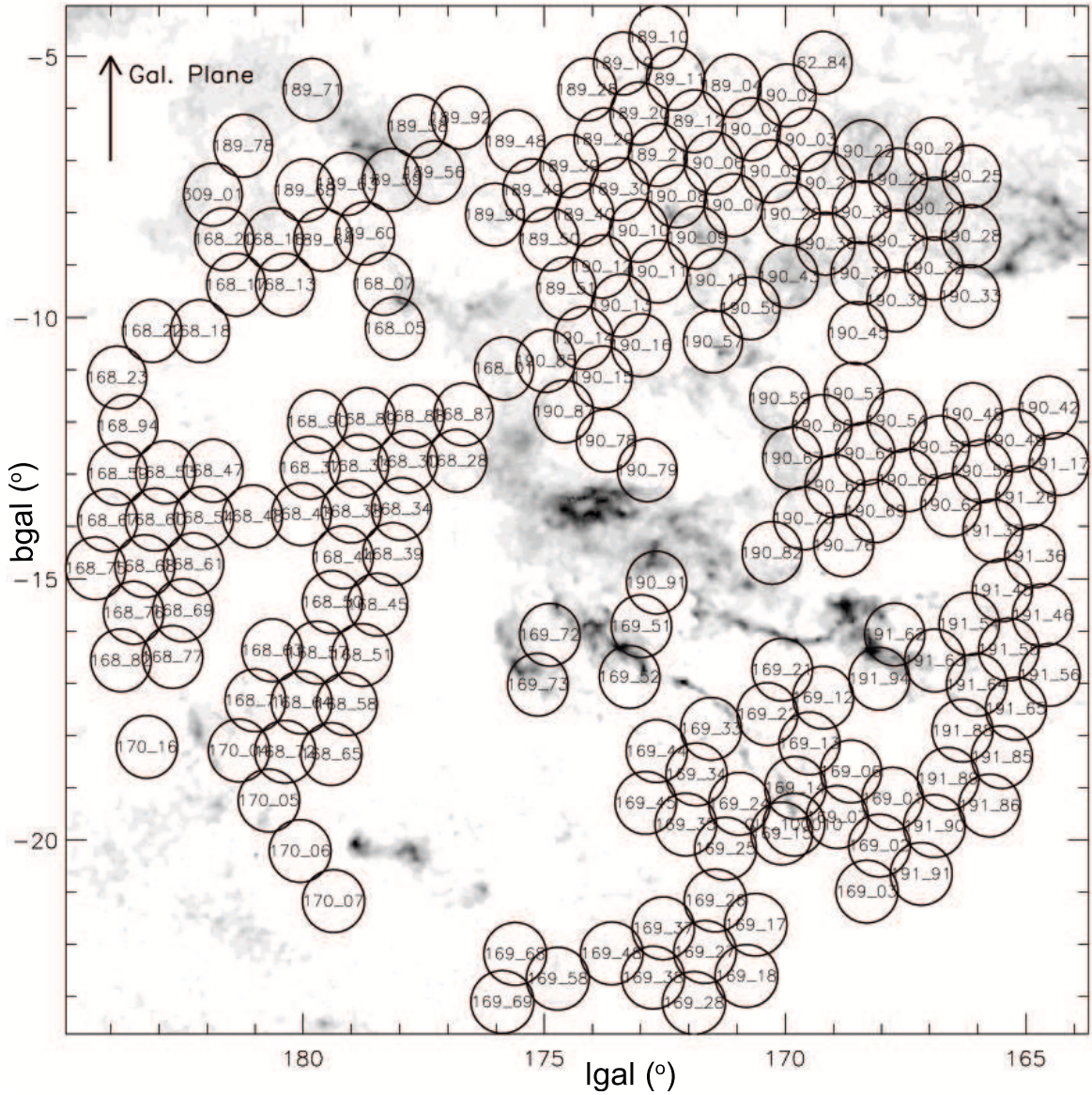}
\caption{Location and tile identification of the GALEX AIS images used in the work. 
The molecular gas distribution, as unveiled by the 2MASS extinction map (Lombardi et al. 2010) is overlaid for reference.  }
\label{ais}
\end{figure}

\newpage

\begin{figure}[tb]
\includegraphics[width=17cm]{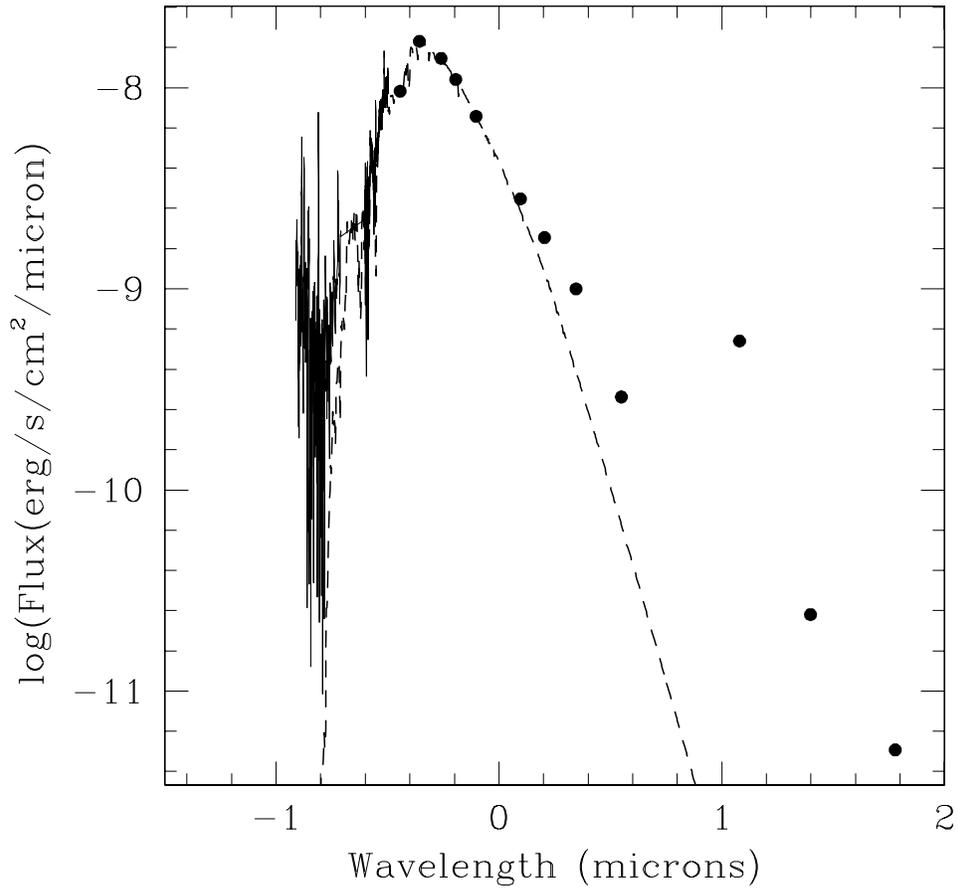}
\caption{Spectral energy distribution of the CTTSs binary AK~Sco composed by two F5 stars. The IUE UV spectrum of AK~Sco as well as the photometric data from Alencar et al. (2003) are represented. The Kurucz model of a F5 star is overlaid (dashed line).}
\label{UVexcess}
\end{figure}

\newpage

\begin{figure}[tb]
\includegraphics[width=17cm]{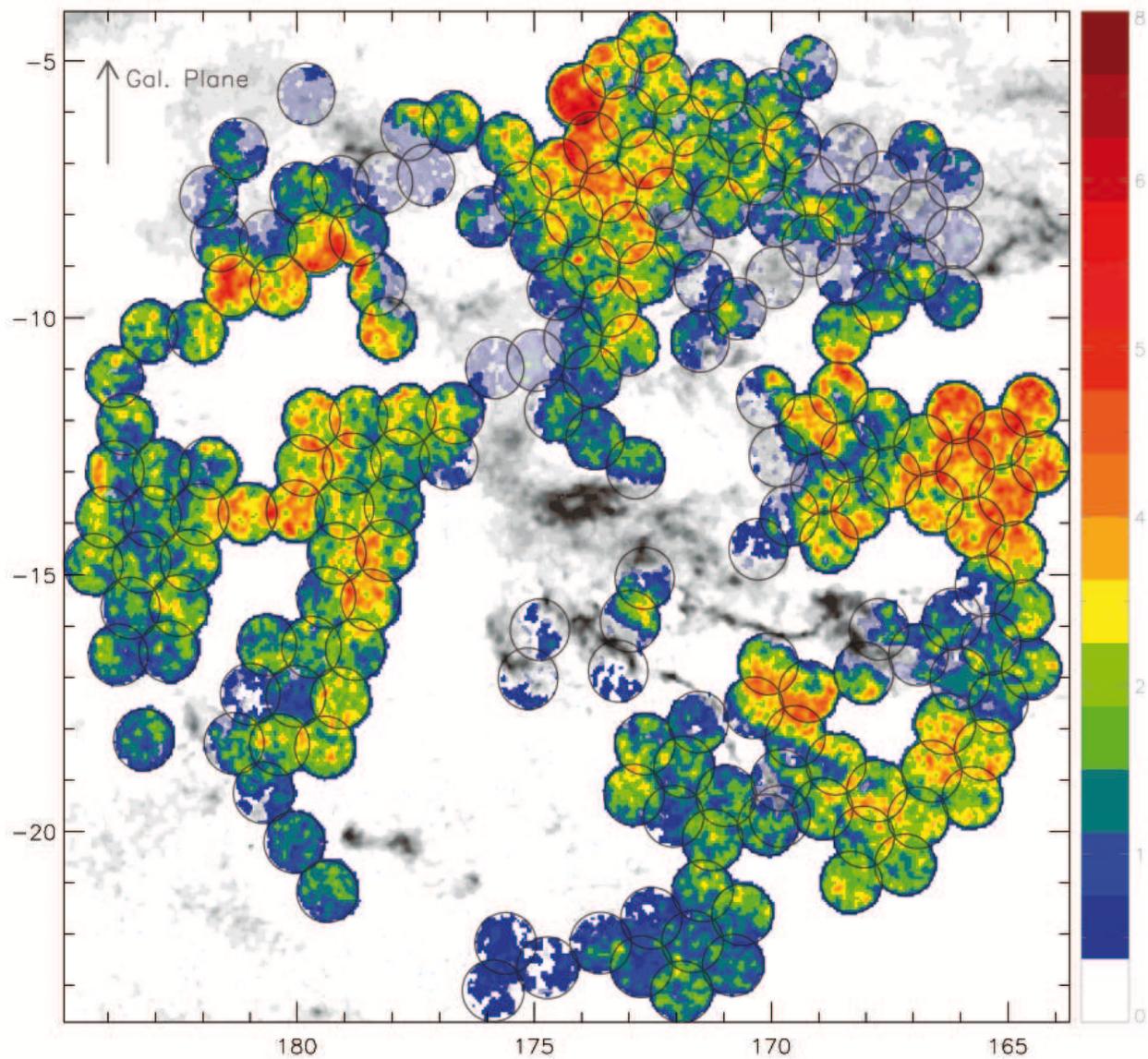}
\caption{Density of NUV GALEX sources in the TMC. The densities are color coded  in stars per 3 arcmin$^2$ (see lateral bar). 
The molecular gas distribution, as unveiled by the 2MASS extinction map (Lombardi et al. 2010) is overlaid for reference.  }
\label{NUV_clean}
\end{figure}

\newpage

\begin{figure}[tb]
\includegraphics[width=17cm]{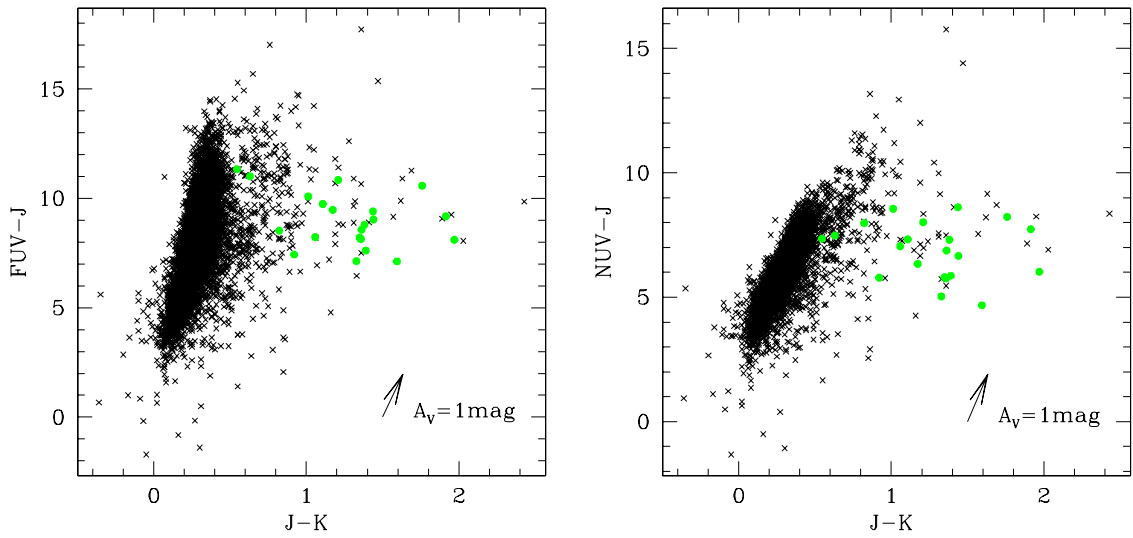}
\caption{Color-color diagrams for the survey stars (black dots) and the IUE TTSs qualification sample (green circles)  to be compared with Findeisen \& Hillenbrand 2010
work.}
\label{Hillen}
\end{figure}

\newpage

\begin{figure}[h]
\includegraphics[width=16cm]{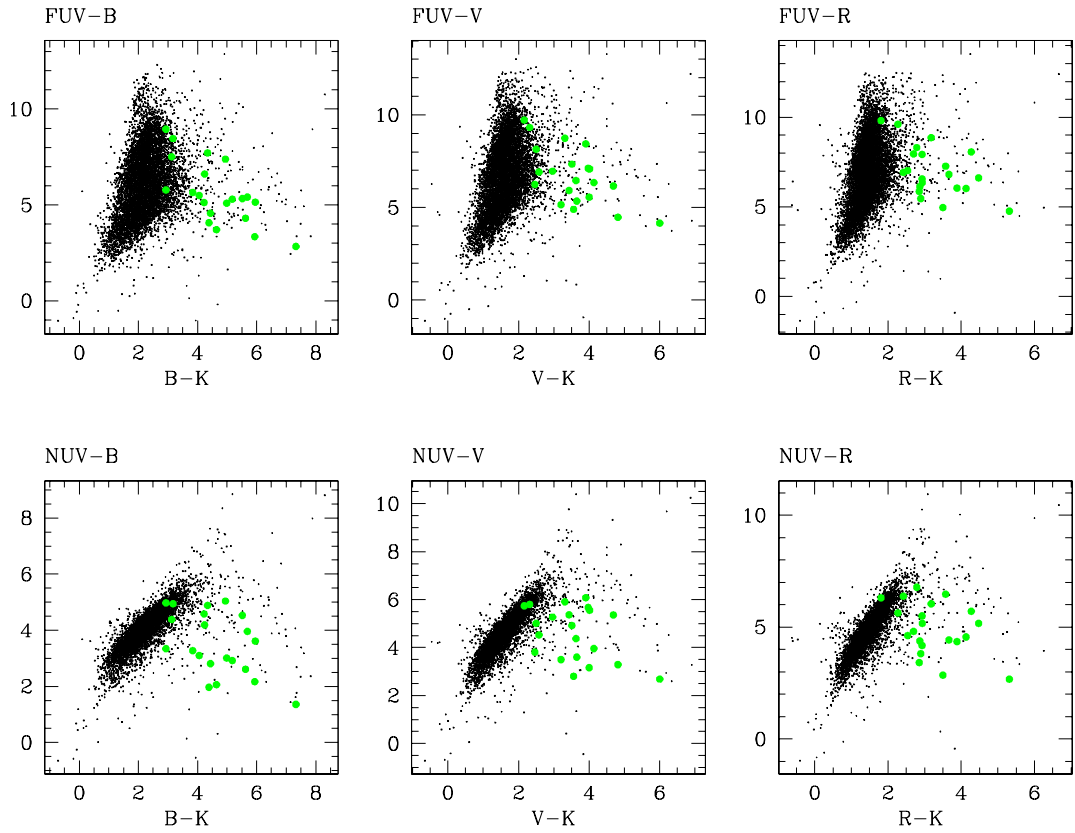}
\caption{Color-color diagrams showing the  UV and infrared excess with respect to the photosphere for the survey stars  (black dots) and the IUE TTSs qualification sample (green circles). }
\end{figure}

\newpage

\begin{figure}[tb]
\includegraphics[width=16cm]{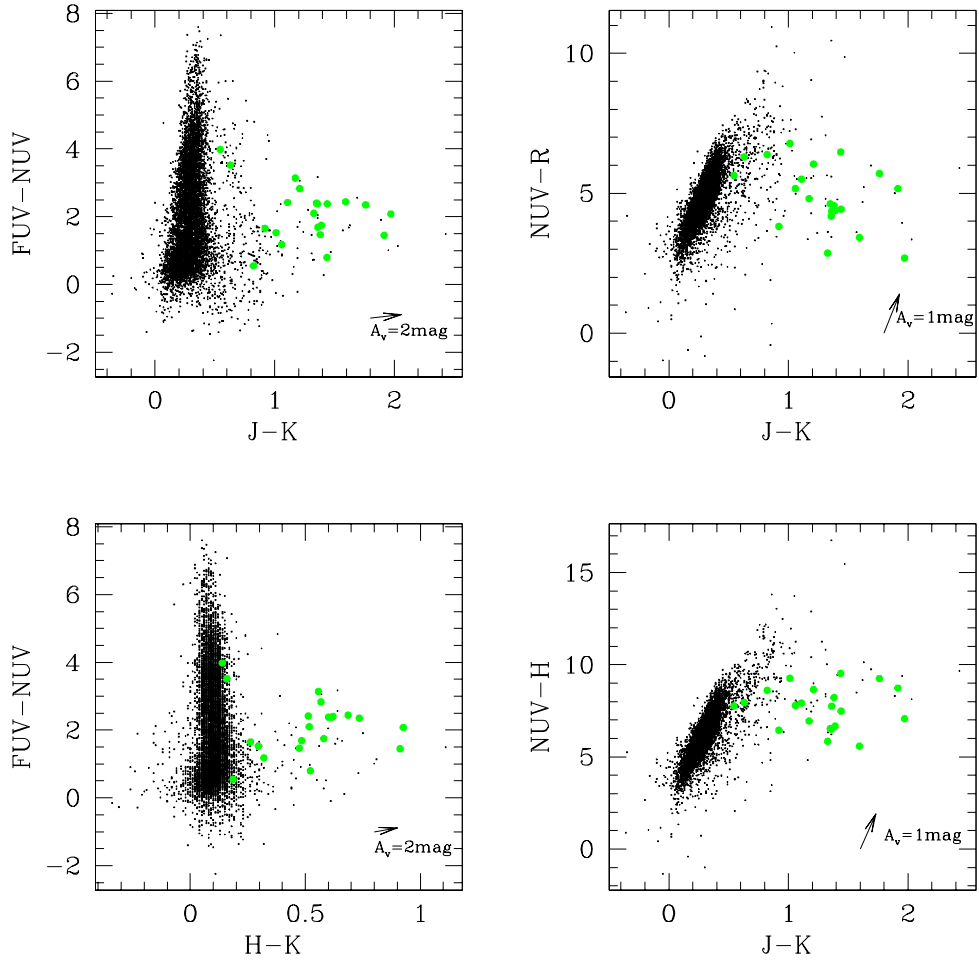}
\caption{Color-color  diagrams used to select candidates to TTSs  in the GALEX fields; survey stars and the IUE qualification sample are 
represented by black dots and green circles, respectively.}
\label{GALEX-TTS}
\end{figure}

\newpage

\begin{figure}[h]
\begin{tabular}{cc}
\includegraphics[width=10cm]{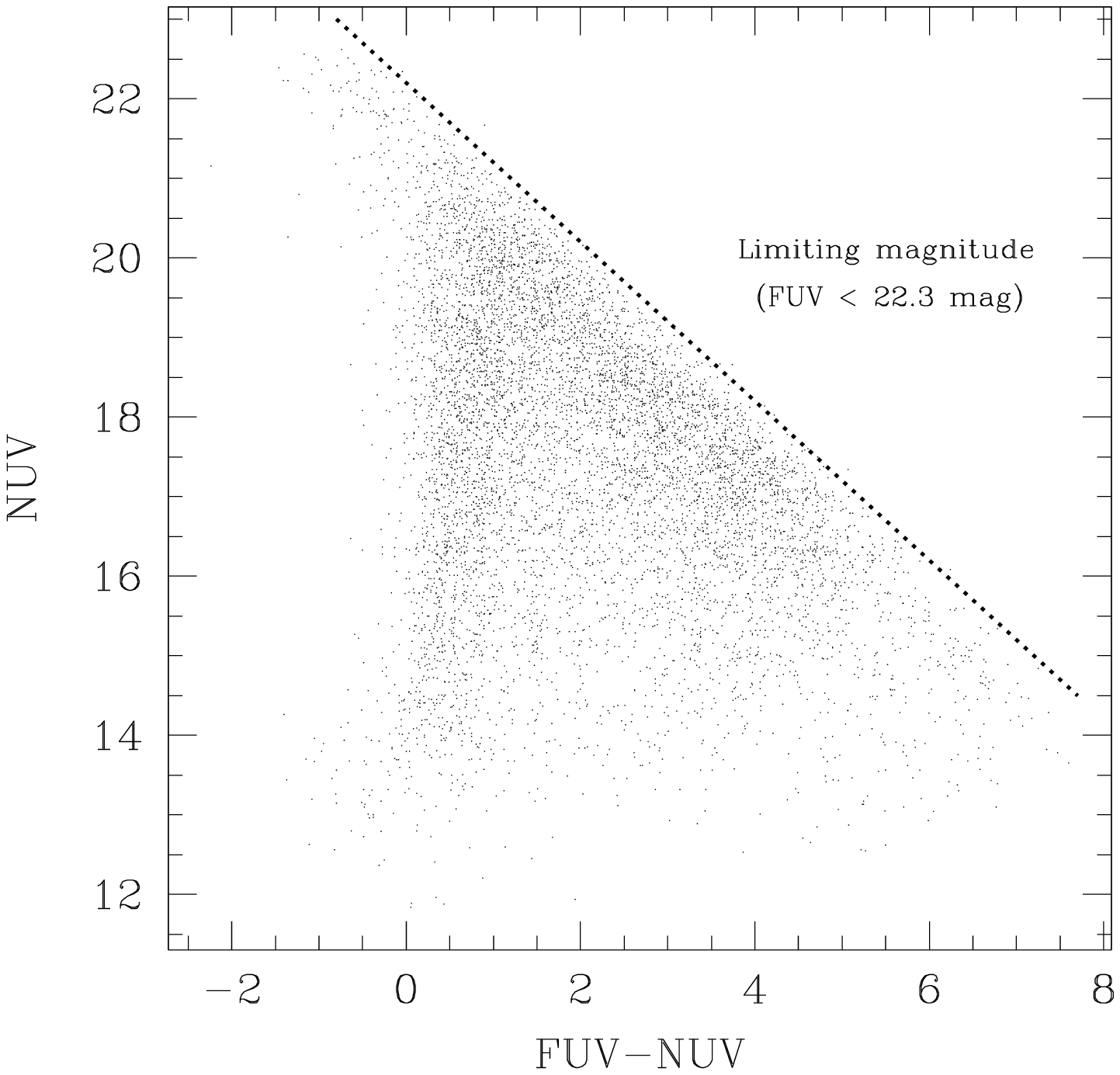}&\includegraphics[width=8cm]{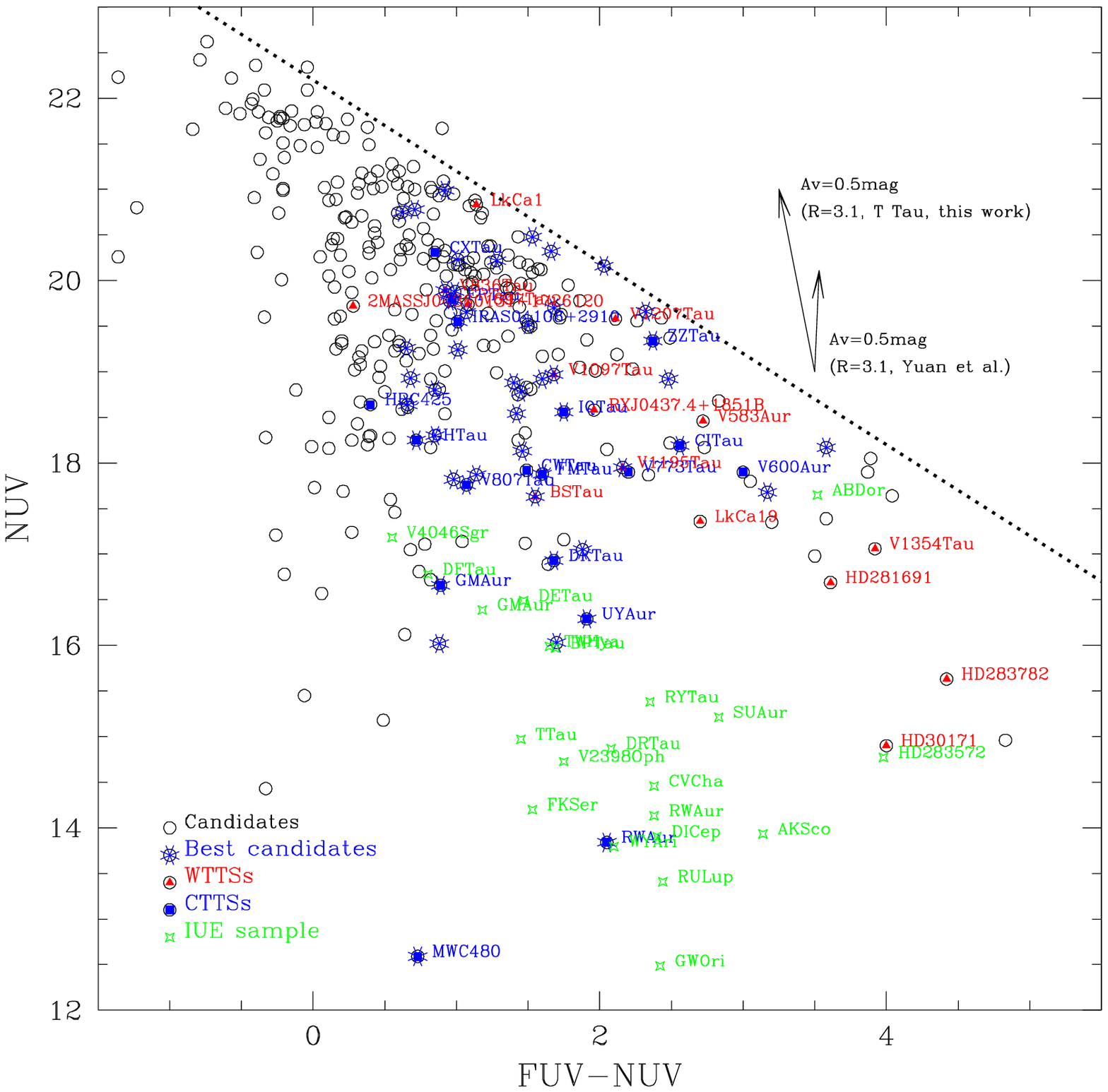}\\
\end{tabular}
\caption{NUV versus FUV-NUV color-magnitude diagram for the FUV sources in the GALEX AIS survey of the TMC. 
{\it Left panel}: all survey sources are represented. The limiting magnitude boundary is marked with a
dashed line. {\it Right panel}:  The candidates found in this work and the IUE sample of TTSs are marked with open black circles and 
green stars, respectively. The magnitudes of the TTSs in the IUE sample have been set at a distance of 140~pc (the distance to the TMC) for reference. 
The candidates satisfying all the four criteria
(see Sect.~3.1) are marked with blue asterisks. The known TTSs recovered in the search for
candidates are indicated: CTTS (blue squares), WTTS (red triangles). Extinction is marked
from Yuan et al. 2013 (for the Fitzpatrick extinction law) and for T Tau; both for R=3.1.}
\label{sel}
\end{figure}

\newpage

\begin{figure}[tb]
\includegraphics[width=16cm]{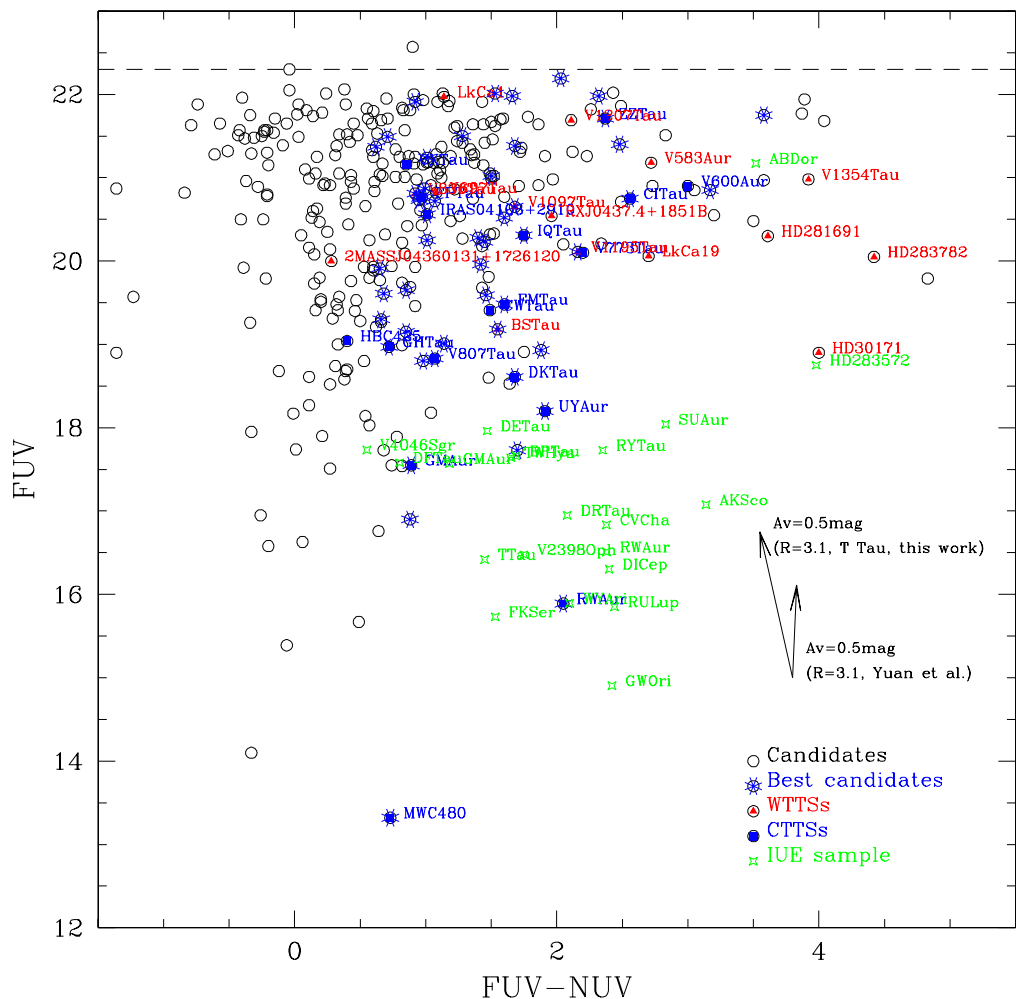}
\caption{As Fig.~9 for the FUV versus FUV-NUV color-magnitude diagram. 
The dashed line represents the FUV magnitude that the 2M1207 brown dwarf would have 
if located at the center of the Hyades open cluster (45 pc).
\label{hist2}}
\end{figure}

\newpage

\begin{figure}[t]
\centering
\includegraphics[width=16cm]{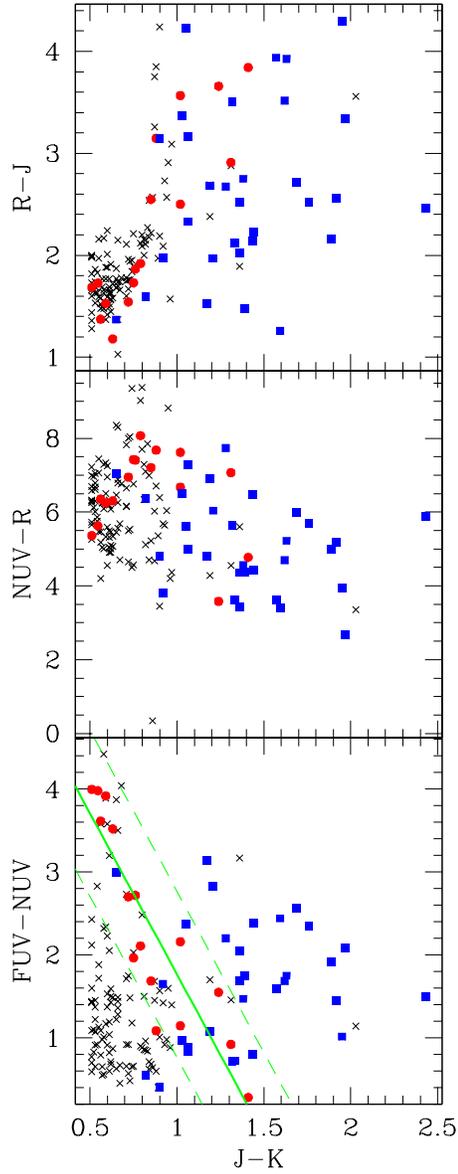}
\caption{Color-color color diagrams for the final selection of TTSs candidates; candidates are marked with black crosses, 
and CTTSs and WTTSs  from the qualification sample are  represented by blue squares and red circles, respectively. The regresion
line of the WTTSs in the FUV-NUV vs J-K diagram is plotted (solid line).}
\label{sed}
\end{figure}

\newpage

\begin{figure}[tb]
\includegraphics[width=16cm]{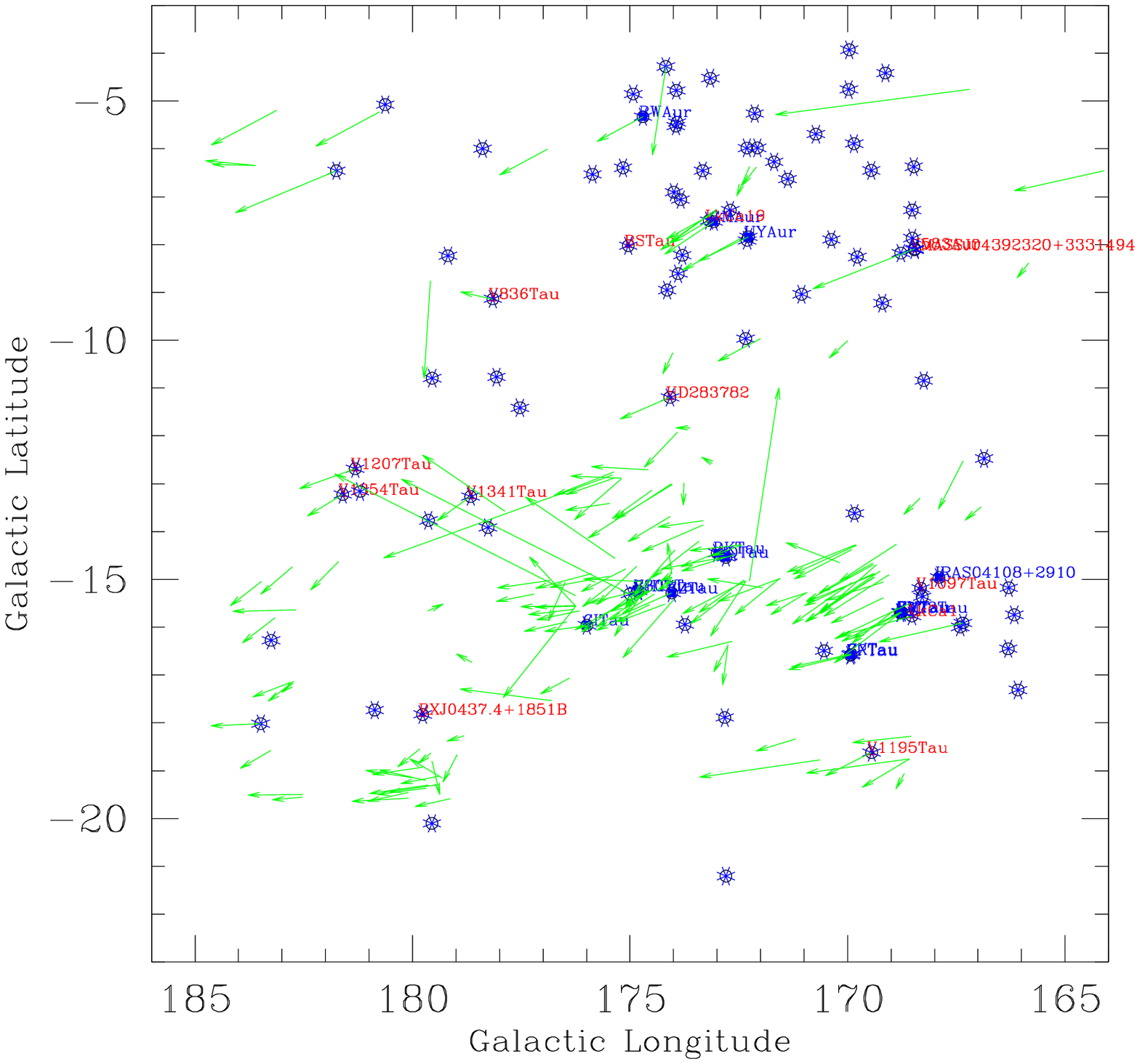}\\
\caption{Location of the TTSs candidates over the TMC. Also the  proper motions of the PMS 
stars in the TMC are indicated (from Doucorant et al., 2005).
The name of the known TTSs are indicated in red for WTTSs and in blue for CTTSs. 
\label{location2}}
\end{figure}

\begin{table}
\caption{T Tauri Stars used as reference sample\tablenotemark{(a,b)}
\label{ttauris}}
\begin{scriptsize}
\begin{tabular}{llclllllll}
\hline \hline
HBC & Star & Spectral\tablenotemark{(c)} & d & FUV & NUV & R & J & H & K \\
    &      &  Type    & pc & AB mag & AB mag &  mag & mag & mag & mag \\
\hline
10 & WY Ari & K5 Bin 	& 275 & 17.36 & 15.26 & 12.4 & 10.229$\pm$0.021 & 9.418$\pm$0.021 & 8.901$\pm$ 0.018 \\ 
32 & BP Tau & K7     	& 140 & 17.66 & 15.97 & 11.62 & 9.098$\pm$0.037 & 8.220$\pm$0.024 & 7.736 $\pm$ 0.023 \\
33 & DE Tau & M2     	& 140 & 17.96 & 16.49 & 11.93 & 9.180$\pm$0.022 & 8.273$\pm$0.018 & 7.799$\pm$0.018 \\
34 & RY Tau & K1     	& 140 & 17.73 & 15.38 &  9.67 & 7.155$\pm$0.019 & 6.13 $\pm$ 0.06 & 5.395$\pm$0.023 \\
35 & T Tau  & K0     	& 140 & 16.42 & 14.97 &  9.8  & 7.240$\pm$0.023 & 6.237$\pm$0.017 & 5.325$\pm$0.017 \\
36 & DF Tau & M0-1   	& 140 & 17.58 & 16.78 & 11.50 & 8.171$\pm$0.026 & 7.256$\pm$0.023 & 6.734$\pm$0.024 \\
74 & DR Tau & K4     	& 140 & 16.95 & 14.87 & 12.19$\pm$0.41 & 8.845$\pm$0.024 & 7.80$\pm$ 0.05 & 6.874$\pm$0.017 \\
77 & GM Aur & K3     	& 140 & 17.57 & 16.39 & 11.22 & 9.341$\pm$0.018 & 8.603$\pm$0.024 &8.283$\pm$0.017 \\
79 & SU Aur & G2     	& 140 & 18.04 & 15.21 & 9.17 & 7.199$\pm$0.020 & 6.558$\pm$0.020 & 5.990$\pm$0.023 \\
80 & RW Aur & K1	& 140 & 16.51 & 14.13 & 9.95 & 8.378$\pm$0.024 & 7.621$\pm$0.038 & 7.020 $\pm$ 0.018 \\
85 & GW Ori & G5 	& 450 & 17.44 & 15.02 & 9.52 & 7.698$\pm$0.030 & 7.103$\pm$ 0.029 & 6.590$\pm$0.029 \\
247 & CV Cha & G8 	& 175 & 17.32 & 14.94 & 10.51 & 8.285$\pm$0.023 & 7.46$\pm$0.04 & 6.845$\pm$0.026 \\
251 & RU Lup & K7 	& 140 & 15.85 & 13.41 & 9.99 & 8.732 $\pm$ 0.026 & 7.824 $\pm$ 0.042 & 7.138$\pm$0.024 \\
271 & AK Sco & F5 SB	& 145 & 17.15 & 14.01 & 9.2 & 7.676$\pm$ 0.026 & 7.06$\pm$0.03 & 6.503$\pm$ 0.020 \\
315 & DI Cep & G8 	& 244 & 17.51 & 15.11 & 10.49 & 9.302$\pm$0.026 & 8.572 $\pm$ 0.047 & 7.952$\pm$ 0.026 \\ 
380 & HD 283572 & G5 	& 140 & 18.75 & 14.77 & 9.14 & 7.414 $\pm$ 0.029 & 7.008$\pm$0.026 & 6.869$\pm$0.023 \\
435 & AB Dor & K0 	&  15 & 16.32 & 12.8  & 6.496 & 5.316 $\pm$ 0.019 & 4.845 $\pm$ 0.033 & 4.686 $\pm$ 0.016 \\
568 & TW Hya & K7 	&  56 & 15.65 & 14.00 & 10.19 & 8.217 $\pm$ 0.024 & 7.558 $\pm$0.042 & 7.297$\pm$ 0.024 \\ 
656 & V2398 Oph & G8 	& 125 & 16.23 & 14.48 & 10.1\tablenotemark{(d)} & 8.62 $\pm$ 0.024 &  7.810 $\pm$ 0.046 & 7.23 $\pm$ 0.08 \\
662 &V4046 Sgr& K5,6 SB &  83 & 16.60 & 16.05 & 9.67\tablenotemark{(e)} & 8.071 $\pm$ 0.023 & 7.435 $\pm$ 0.051 & 7.249 $\pm$ 0.020 \\
664 & FK Ser & K7 Bin	& 350 & 17.72 & 16.19 & 9.41 & 7.636$\pm$0.020 & 6.92 $\pm$0.03 & 6.624$\pm$0.021 \\ 
\hline
\end{tabular}
\tablenotetext{(a)}{see G\'omez de Castro \& Franqueira 1997 for more details and R values}
\tablenotetext{(b)}{J H K mag from 2MASS}
\tablenotetext{(c)}{SB: Spectroscopic Binary}
\tablenotetext{(d)}{UCAC4 catalogue, Zacharias et al. 2013} 
\tablenotetext{(e)}{Hutchinson et al. 1990}
\end{scriptsize}
\end{table}

\newpage
\begin{table}
\caption{Extract of the table with the TTSs and candidates found in the TMC from the GALEX AIS$^{(a)}$}
\begin{scriptsize}
\begin{tabular}{llllllllll}
\hline \hline
RA(deg)  & DEC(deg) & FUV(mag)	& NUV(mag) & R2(mag) & J(mag)	& H(mag) & K(mag) &  ID  & Obj. Type \\

\hline
60.18 & 28.87 & 20.21 & 17.87 & 12.54 & 11.17 & 10.66 & 10.58 &     &Candidate \\
61.04 & 29.34 & 21.77 &  17.9 & 11.44 &  9.57 &  9.05 &  8.92 &     &Candidate WTTS \\
61.5 & 29.94 & 20.85 & 17.68 & 12.08 & 10.19 &  9.47 &  8.83 &     &Candidate CTTS \\
61.71 & 25.69 & 20.11 & 17.95 & 11.27 &  8.77 &  8.03 &  7.75   & V1195Tau &WTTS \\
62.06 & 21.63 & 21.81 & 20.93 & 11.55 &  9.46 &  8.84 &  8.66 &     &Candidate \\
62.06 & 30.26 & 21.37 & 20.75 & 16.04 & 13.86 & 13.31 & 13.02 &     &Candidate \\
62.27 & 28.92 & 21.75 & 18.17 & 11.97 &  9.96 &  9.57 &  9.36 &     &Candidate WTTS \\
62.29 & 29.02 &  20.3 & 16.69 & 10.33 &  8.96 &  8.51 &   8.4   & HD281691 &Candidate WTTS$^{(b)}$ \\
\hline
\end{tabular}
\begin{tabular}{l}
(a): Full table available in electronic format.\\
(b): Also candidate to TTSs from the proper motions survey by Doucorant et al. (2005). \\
\end{tabular}
\end{scriptsize}
\end{table}

\end{document}